\documentclass[aps,tightenlines]{revtex4}
\usepackage{graphicx}
\usepackage{float}
\begin{document}
\title{Non--linear fractal interpolating functions of
one and two variables}
\author{ R.~Kobes}
\affiliation{Department of Physics and Winnipeg Institute for 
Theoretical Physics, \\University of Winnipeg, Winnipeg, 
Manitoba, R3B 2E9 Canada}
\email{r.kobes@uwinnipeg.ca}
\author{ A.~J.~Penner}
\affiliation{Department of Physics, University of Manitoba, \\ Winnipeg, 
Manitoba, R2T 2N2 Canada}
\email{ajpenner@physics.umanitoba.ca}
\begin{abstract}
We consider non--linear generalizations of fractal
interpolating functions applied to functions
of one and two variables.
The use of such interpolating functions
in resizing images is illustrated.
\end{abstract}
\maketitle
\section{Introduction}
\par An Iterated Function System (IFS) may be used to
construct fractal interpolation functions for some
data \cite{barn,ifs,sv}. 
The simplest example of interpolating a function
$x(t)$, given data points ($t_i, x_i$), $i=0,1,\ldots, N$,
starts with an IFS
\begin{equation}
W_n\left(\begin{array}{c}t\\x\end{array}\right) =
\left( \begin{array}{cc} a_n & 0\\ c_n & 0\end{array} \right)
\left(\begin{array}{c}t\\x\end{array}\right) +
\left(\begin{array}{c}e_n\\f_n\end{array}\right).
\label{onevar}
\end{equation}
The coefficients $a_n, c_n, e_n$, and $f_n$ determined
from the conditions, for $n=1,2,\ldots,N$,
\begin{equation}
W_n\left(\begin{array}{c}t_0\\x_0\end{array}\right) =
\left(\begin{array}{c}t_{n-1}\\x_{n-1}\end{array}\right),
\qquad
W_n\left(\begin{array}{c}t_N\\x_N\end{array}\right) =
\left(\begin{array}{c}t_{n}\\x_{n}\end{array}\right).
\label{lincond}
\end{equation}
which leads to
\begin{equation}
a_n = \frac{t_n - t_{n-1}}{t_N-t_0}, \quad
e_n =  \frac{t_{n-1}t_N - t_{n}t_0}{t_N-t_0} \quad
c_n =  \frac{x_n - x_{n-1}}{t_N-t_0}, \quad
f_n =  \frac{x_{n-1}t_N - x_nt_{0}}{t_N-t_0},
\label{lincoeff}
\end{equation}
With these, the transformation of Eq.~(\ref{onevar}) can be written as
\begin{eqnarray}
W_n(t) \equiv t^\prime &=&
\frac{(t-t_0)}{(t_N-t_0)}\ t_n
+\frac{(t-t_N)}{(t_0-t_N)}\ t_{n-1}
\nonumber\\
W_n(x) \equiv x^\prime &=&
\frac{(t^\prime-t_{n-1})}{(t_n-t_{n-1})}\ x_n
+\frac{(t^\prime-t_n)}{(t_{n-1}-t_n)}\ x_{n-1}
\label{linint}
\end{eqnarray}
in which form it is apparent
$W_n(x)\equiv x^\prime$ is determined by a linear
(in $t$) interpolating function
between the points ($t_{n-1}, x_{n-1}$)
and ($t_n, x_n$). Graphs of fractal interpolating
functions can then be made by applying the random iteration
algorithm:
\begin{itemize}
\item {\tt initialize} ($t, x$) {\tt to a point in the interval of interest}
\item {\tt for a set number of iterations}
  \begin{itemize}
  \item {\tt randomly select a transformation} $W_n(t, x)$
  \item {\tt plot} ($t^\prime, x^\prime $) $ = W_n(t, x)$
  \item {\tt set} ($t, x$) $ = $ ($t^\prime, x^\prime $)
  \end{itemize}
 \item {\tt end for}
\end{itemize}
\par
In this paper we consider two non--linear generalizations
of such fractal interpolating functions. The first 
concerns how to extend the linear interpolation of
Eq.~(\ref{linint}) to higher--degree interpolations.
The second generalization arises when one considers
the construction of fractal interpolating functions
for functions of two (or more) variables -- in this case, even
a linear interpolation of the form of Eq.~(\ref{linint}),
when applied to each variable,
will result in a non--linear interpolating function. This has 
an obvious application in
representing two--dimensional images, 
where a function of two variables (the pixel coordinates)
could represent a black--and--white image (using a Boolean
function), a gray--scale image (using a scalar function),
or a colour image (using a vector--valued
function of the three {\it rgb} [red, green, blue] values).
This problem has been examined extensively in the
context of image compression \cite{jaq,comp1,comp2,mass};
in the last section we consider a related problem
of using these iterated function systems to rescale
images, or portions thereof.
\section{Functions of two variables}
We first consider a function $z(x,y)$ of two variables,
and examine the problem of constructing a fractal
interpolating function from the data $x_i, y_j, z_{i,j}$,
where $ i = 0,1,\ldots, M$, $j=0, 1, \ldots, N$,
 and $z_{i,j}\equiv z(x_i, y_j)$.
To this end, consider the transformation
\begin{eqnarray}
W_{mn}(x) &=& a_{mn} x+e_{mn}\nonumber\\
W_{mn}(y) &=& c_{mn} y+f_{mn}\nonumber\\
W_{mn}(z) &=& A_{mn} x + B_{mn} y +C_{mn} xy +D_{mn} \\
\label{twovar}\end{eqnarray}
We then impose, for $m = 1,2,\ldots,M$ and
$n=1,2,\ldots,N$  the conditions
\begin{eqnarray}
& &W_{mn}\left(\begin{array}{c}x_M\\y_N\end{array}\right) =
\left(\begin{array}{c}x_{m}\\y_{n}\end{array}\right),
\qquad
W_{mn}\left(\begin{array}{c}x_0\\y_0\end{array}\right) =
\left(\begin{array}{c}x_{m-1}\\y_{n-1}\end{array}\right),
\nonumber\\
& &W_{mn}(z_{M,N}) = z_{m,n}, \qquad
W_{mn}(z_{M,0}) = z_{m,n-1}, \qquad
W_{mn}(z_{0,N}) = z_{m-1,n}, \qquad
W_{mn}(z_{0,0}) = z_{m-1,n-1}.
\end{eqnarray}
The coefficients turn out to be
\begin{eqnarray}
a_{mn} &=& \frac{x_{m}-x_{m-1}}{x_{M}-x_{0}},
\qquad
e_{mn} = \frac{x_{0}x_{m}-x_{m-1}x_{M}}{x_{0}-x_{M}},
\nonumber\\
 c_{mn} &=& \frac{y_{n}-y_{n-1}}{y_{N}-y_{0}},
\qquad 
 f_{mn} = \frac{y_{0}y_{n}-y_{n-1}y_{N}}{y_{0}-y_{N}},
\nonumber\\
 A_{mn} &=&
\frac{(z_{m,n-1}-z_{m-1,n-1})y_{N}-(z_{m,n}-z_{m-1,n})y_{0}}
{(x_{M}-x_{0})(y_{N}-y_{0})},
\nonumber\\
 B_{mn} &=&
\frac{(z_{m-1,n}-z_{m-1,n-1})x_{M}-(z_{m,n}-z_{m,n-1})x_{0}}
{(x_{M}-x_{0})(y_{N}-y_{0})},
\nonumber\\
 C_{mn} &=& 
\frac{z_{m,n}-z_{m,n-1}-z_{m-1,n}+z_{m-1,n-1}}
{(x_{M}-x_{0})(y_{N}-y_{0})},
\nonumber\\
 D_{mn} &=&
\frac{z_{m,n}x_0y_0-z_{m,n-1}x_0y_N-z_{m-1,n}x_Ny_0+z_{m-1,n-1}x_Ny_N}
{(x_{M}-x_{0})(y_{N}-y_{0})}.
\end{eqnarray}
With these, the transformation of Eq.~(\ref{twovar}) can be written as
\begin{eqnarray}
W_{mn}(x) \equiv x^\prime &=&
\frac{(x-x_0)}{(x_M-x_0)}\ x_m
+\frac{(x-x_M)}{(x_0-x_M)}\ x_{m-1}
\nonumber\\
W_{mn}(y) \equiv y^\prime &=&
\frac{(y-y_0)}{(y_N-y_0)}\ y_n
+\frac{(y-y_N)}{(y_0-y_N)}\ y_{n-1}
\nonumber\\
W_{mn}(z) \equiv z^\prime &=&
\frac{(x^\prime-x_{m-1})(y^\prime-y_{n-1})}
{(x_m-x_{m-1})(y_n-y_{n-1})}\ z_{m,n} +
\frac{(x^\prime-x_{m})(y^\prime-y_{n-1})}
{(x_{m-1}-x_{m})(y_{n}-y_{n-1})}\ z_{m-1,n} +
\nonumber\\ \qquad &+&
\frac{(x^\prime-x_{m-1})(y^\prime-y_{n})}
{(x_{m}-x_{m-1})(y_{n-1}-y_{n})}\ z_{m,n-1} +
\frac{(x^\prime-x_{m})(y^\prime-y_{n})}
{(x_{m-1}-x_{m})(y_{n-1}-y_{n})}\ z_{m-1,n-1}
\end{eqnarray}
in which form it is apparent
$W_{mn}(z)\equiv z^\prime$ is determined by a function
implementing a linear interpolation over the grid
($x_{m-1}, y_{n-1}$), ($x_{m-1}, y_{n}$),
($x_{m}, y_{n-1}$), and ($x_m, y_n$).
\section{Quadratic interpolating functions}
The interpolating functions considered up to now have used
a linear interpolating formula between adjacent points to
construct the IFS. In this section we indicate how this
can be generalized to quadratic interpolations.
\subsection{Functions of one variable}
For a function $x(t)$ of one variable, using data points
($t_i, x_i$), $i=0,1,\ldots, N$, consider the transformations
\begin{eqnarray}
W_n(t) &=& a_n t + e_n\nonumber\\
W_n(x) &=& c_nt +d_nt^2 +f_n
\end{eqnarray}
and impose the conditions, for $n=2,3,\ldots,N$,
\begin{equation}
W_n\left(\begin{array}{c}t_0\\x_0\end{array}\right) =
\left(\begin{array}{c}t_{n-2}\\x_{n-2}\end{array}\right),
\qquad
W_n\left(\begin{array}{c}t_M\\x_M\end{array}\right) =
\left(\begin{array}{c}t_{n-1}\\x_{n-1}\end{array}\right),
\qquad
W_n\left(\begin{array}{c}t_N\\x_N\end{array}\right) =
\left(\begin{array}{c}t_{n}\\x_{n}\end{array}\right).
\label{quadcond}
\end{equation}
The point $t_M$ is determined as
\begin{equation}
t_M = \frac{(t_{n-1}-t_{n-2})}{(t_n-t_{n-2})}\ t_N
+\frac{(t_{n-1}- t_{n})}{(t_{n-2}-t_{n})}\ t_0
\label{tm}
\end{equation}
with corresponding point $x_m$. The coefficients of the IFS are
determined as
\begin{eqnarray}
a_n &=& \frac{t_n-t_{n-2}}{t_N-t_0}\nonumber\\
e_n &=& \frac{t_Nt_{n-2}-t_0t_n}{t_N-t_0}\nonumber\\
c_n &=& \frac{x_n(t_0^2-t_m^2) +x_{n-1}(t_N^2-t_0^2)+x_{n-2}(t_m^2-t_N^2)}
{ (t_N-t_0)(t_N-t_m)(t_m-t_0)} \nonumber\\
d_n &=& \frac{x_n(t_m-t_0) +x_{n-1}(t_0-t_N)+x_{n-2}(t_N-t_m)}
{ (t_N-t_0)(t_N-t_m)(t_m-t_0)} \nonumber\\
f_n &=& \frac{x_nt_mt_0(t_m-t_0) +x_{n-1}t_Nt_0(t_0-t_N)
+x_{n-2}t_Nt_m(t_N-t_m)}
{ (t_N-t_0)(t_N-t_m)(t_m-t_0)}
\end{eqnarray}
with which the transformation can be written as
\begin{eqnarray}
W_n(t) \equiv t^\prime &=&
\frac{(t-t_0)}{(t_N-t_0)}\ t_n
+\frac{(t-t_N)}{(t_0-t_N)}\ t_{n-2}
\nonumber\\
W_n(x) \equiv x^\prime &=&
\frac{(t^\prime-t_{n-1})(t^\prime-t_{n-2})}
{(t_n-t_{n-1})(t_n-t_{n-2})}\ x_n
+\frac{(t^\prime-t_{n})(t^\prime-t_{n-2})}
{(t_{n-1}-t_{n})(t_{n-1}-t_{n-2})}\ x_{n-1}\nonumber\\
&+&\frac{(t^\prime-t_{n-1})(t^\prime-t_{n})}
{(t_{n-2}-t_{n-1})(t_{n-2}-t_{n})}\ x_{n-2}
\label{quadint}
\end{eqnarray}
In this form we see that a quadratic (in $t^\prime$) interpolating
function is used between the points ($t_n, x_n$), ($t_{n-1}, x_{n-1}$),
and ($t_{n-2}, x_{n-2}$).
\subsection{Functions of two variables}
We next consider a function $z(x,y)$ of two variables,
and construct a fractal interpolating function which employs
a quadratic interpolation between points. 
To this end, consider the transformation
\begin{eqnarray}
W_{mn}(x) &=& a_{mn} x+e_{mn}\nonumber\\
W_{mn}(y) &=& c_{mn} y+f_{mn}\nonumber\\
W_{mn}(z) &=& A_{mn} x^2y^2 + B_{mn} x^2y +C_{mn} x^2 +D_{mn}xy^2 + E_{mn} xy + F_{mn}x
+G_{mn}y^2+H_{mn}y+I_{mn}
\label{twovarq}\end{eqnarray}
We then impose, for $m =2,3,\ldots,M$ and $n=2,3,\ldots,N$, the conditions
\begin{eqnarray}
& &W_{mn}\left(\begin{array}{c}x_M\\y_N\end{array}\right) =
\left(\begin{array}{c}x_{m}\\y_{n}\end{array}\right),
\qquad
W_{mn}\left(\begin{array}{c}x_M\\y_M\end{array}\right) =
\left(\begin{array}{c}x_{m-1}\\y_{n-1}\end{array}\right),
\qquad
W_{mn}\left(\begin{array}{c}x_0\\y_0\end{array}\right) =
\left(\begin{array}{c}x_{m-2}\\y_{n-2}\end{array}\right),
\nonumber\\
& &W_{mn}(z_{N,N}) = z_{m,n}, \qquad
W_{mn}(z_{N,m}) = z_{m,n-1}, \qquad
W_{mn}(z_{N,0}) = z_{m,n-2}, \nonumber\\
& &W_{mn}(z_{m,N}) = z_{m-1,n}, \qquad
W_{mn}(z_{m,m}) = z_{m-1,n-1} \qquad
W_{mn}(z_{m,0}) = z_{m-1,n-2}\nonumber\\
& &W_{mn}(z_{0,N}) = z_{m-2,n}, \qquad
W_{mn}(z_{0,m}) = z_{m-2,n-1} \qquad
W_{mn}(z_{0,0}) = z_{m-2,n-2}
\label{cond2}\end{eqnarray}
The points $x_M$ and $y_M$ are determined as
\begin{eqnarray}
x_M &=& \frac{(x_{m-1}-x_{m-2})}{(x_m-x_{m-2})}\ x_M
+\frac{(x_{m-1}- x_{m})}{(x_{m-2}-x_{m})}\ x_0 \nonumber\\
y_M &=& \frac{(y_{n-1}-y_{n-2})}{(y_n-y_{n-2})}\ y_N
+\frac{(y_{n-1}- y_{n})}{(y_{n-2}-y_{n})}\ y_0,
\label{tm2}
\end{eqnarray}
along with the corresponding $z$ points.
The coefficients of the IFS can then be determined, 
by which the transformation of Eq.~(\ref{cond2}) can be
written as
\begin{eqnarray}
W_{mn}(x) \equiv x^\prime &=&
\frac{(x-x_0)}{(x_M-x_0)}\ x_m
+\frac{(x-x_M)}{(x_0-x_M)}\ x_{m-1}
\nonumber\\
W_{mn}(y) \equiv y^\prime &=&
\frac{(y-y_0)}{(y_N-y_0)}\ y_n
+\frac{(y-y_N)}{(y_0-y_N)}\ y_{n-1}
\nonumber\\
W_{mn}(z) \equiv z^\prime &=&
\frac{ 
(x^\prime - x_{m-2})(x^\prime - x_{m-1})
(y^\prime - y_{n-2})(y^\prime - y_{n-1})}{
(x_m - x_{m-2}) (x_m-x_{m-1})
(y_n - y_{n-2}) (y_n-y_{n-1}) }z_{m,n} + \nonumber\\
&+& \frac{
(x^\prime - x_{m-2})(x^\prime - x_{m-1})
(y^\prime - y_{n-2})(y^\prime - y_{n})}{
(x_m - x_{m-2}) (x_m-x_{m-1})
(y_{n-1} - y_{n-2}) (y_{n-1}-y_{n}) }z_{m,n-1} + \nonumber\\
&+& \frac{
(x^\prime - x_{m-2})(x^\prime - x_{m-1})
(y^\prime - y_{n})(y^\prime - y_{n-1})}{
(x_m - x_{m-2}) (x_m-x_{m-1})
(y_{n-2} - y_{n}) (y_{n-2}-y_{n-1}) }z_{m,n-2} + \nonumber\\
&+& \frac{
(x^\prime - x_{m})(x^\prime - x_{m-2})
(y^\prime - y_{n-1})(y^\prime - y_{n-2})}{
(x_{m-1} - x_{m}) (x_{m-1}-x_{m-2})
(y_{n} - y_{n-1}) (y_{n}-y_{n-2}) }z_{m-1,n}  + \nonumber\\
&+& \frac{
(x^\prime - x_{m})(x^\prime - x_{m-2})
(y^\prime - y_{n-2})(y^\prime - y_{n})}{
(x_{m-1} - x_{m}) (x_{m-1}-x_{m-2})
(y_{n-1} - y_{n-2}) (y_{n-1}-y_{n}) }z_{m-1,n-1} + \nonumber\\
&+& \frac{
(x^\prime - x_{m})(x^\prime - x_{m-2})
(y^\prime - y_{n})(y^\prime - y_{n-1})}{
(x_{m-1} - x_{m}) (x_{m-1}-x_{m-2})
(y_{n-2} - y_{n}) (y_{n-2}-y_{n-1}) }z_{m-1,n-2}  + \nonumber\\
&+& \frac{
(x^\prime - x_{m})(x^\prime - x_{m-1})
(y^\prime - y_{n-2})(y^\prime - y_{n-1})}{
(x_{m-2} - x_{m}) (x_{m-2}-x_{m-1})
(y_{n} - y_{n-2}) (y_{n}-y_{n-1}) }z_{m-2,n} +  \nonumber\\
&+& \frac{
(x^\prime - x_{m})(x^\prime - x_{m-1})
(y^\prime - y_{n})(y^\prime - y_{n-2})}{
(x_{m-2} - x_{m}) (x_{m-2}-x_{m-1})
(y_{n-1} - y_{n}) (y_{n-1}-y_{n-2}) }z_{m-2,n-1}  + \nonumber\\
&+& \frac{
(x^\prime - x_{m})(x^\prime - x_{m-1})
(y^\prime - y_{n})(y^\prime - y_{n-1})}{
(x_{m-2} - x_{m}) (x_{m-2}-x_{m-1})
(y_{n-2} - y_{n}) (y_{n-2}-y_{n-1}) }z_{m-2,n-2}
\end{eqnarray}
Although tedious to work out, the generalization of the preceding
considerations to higher--order interpolating functions is 
straightforward in principle.
\section{Image Scaling}
As an application of the preceding,
in this section we consider the task of rescaling a
colour image. This is a natural problem for an interpolating
function of two variables ($x, y$) interpreted as pixel
coordinates -- the function ${\bf \vec z}(x, y)$ in this case 
will be a vector--valued function having three components
representing the {\it rgb} value of the pixel specifying the
amount of red, green, and blue present.
\par
The procedure used to scale an image of size $M$ pixels wide by
$N$ pixels high is as follows. We
first read in the rgb values of each pixel of the image, and 
use that as the data to construct a fractal interpolating
function ${\bf \vec z}(i, j)$,
where $i=1,2,\ldots, M$ and $j=1,2,\ldots, N$. 
To then resize the image, so that the resulting
image is of size $s_xM \times s_yN$, we construct a new
fractal interpolating function 
${\bf \vec z}\,^\prime(i,j) = {\bf \vec z}(s_xi,s_yj)$.
Applying the random iteration algorithm to ${\bf \vec z}\,^\prime(i,j)$,
choosing independently a transformation index ($i, j$) at each
stage, will then result in the rescaled image. The generalization
of this procedure to rescale a portion of an image is straightforward.
\par
As examples of the results of this procedure, consider the
figures in the Appendix. We start with the image appearing
in Fig.~\ref{lena}, and zoom in on the area of the face.
The result appears in Fig.~\ref{lena-face}, together with
a comparison done using a simple 
linear interpolation scheme. Zooming further into the area
of the eye results in Fig.~\ref{lena-eye}, again with a
comparison of the result of a simple linear interpolation.
Generally, the number of iterations needed in the random
iteration algorithm to produce acceptable images is of the
order of $s_xM \times s_yN$, where the original image is of
size $M \times N$. Also, while slower, the quadratic fractal
interpolating function typically produces, for the same number
of iterations, a ``smoother'' looking image than the corresponding
linear interpolating function. However, as with all interpolation
schemes, there comes a point where such higher--order interpolating
formulas actually start to produce worse results due to 
an artificially high sensitivity to fluctuations in the data.
\par
Some informal tests of this procedure
seems to indicate that better results are obtained for 
images of people, natural scenery, etc., as opposed to those
containing lettering, simple geometric shapes, and similar
constructs. This might be expected, given the general
fractal nature of such objects in nature. However, as with
all interpolating functions, it is important to remember
that no structural information beyond that of the original
image is being provided (for example, one could not zoom
in on the face of Fig.~\ref{lena} to such a degree as to
see individual skin pores).
\par
The preceding demonstrates that these non--linear
fractal interpolating functions of two variables can
be used in principle to represent images. It would be
interesting to extend these considerations to the
case of partitioned iterated function systems, upon which
much work has been done with respect to compressing
images \cite{jaq}. Work along these directions is in
progress.
\acknowledgments
This work was supported by the Natural Sciences and Engineering
Research Council of Canada.

\appendix
\subsection*{Scaling of figures}
\par\begin{figure}[ht]
\begin{center}
\includegraphics[width=6.5cm]{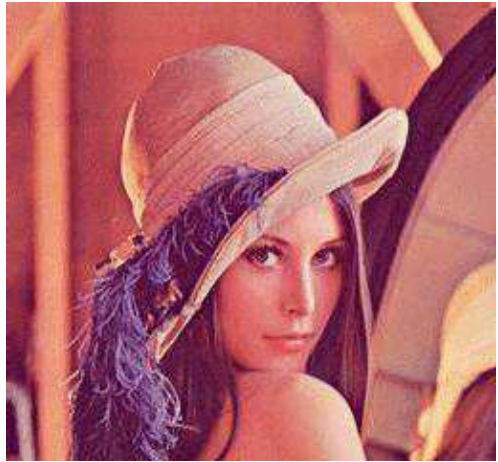}
\end{center}
\caption{Original figure}\label{lena}
\end{figure}
\par\begin{figure}[ht]
\begin{center}
\includegraphics[width=6.5cm]{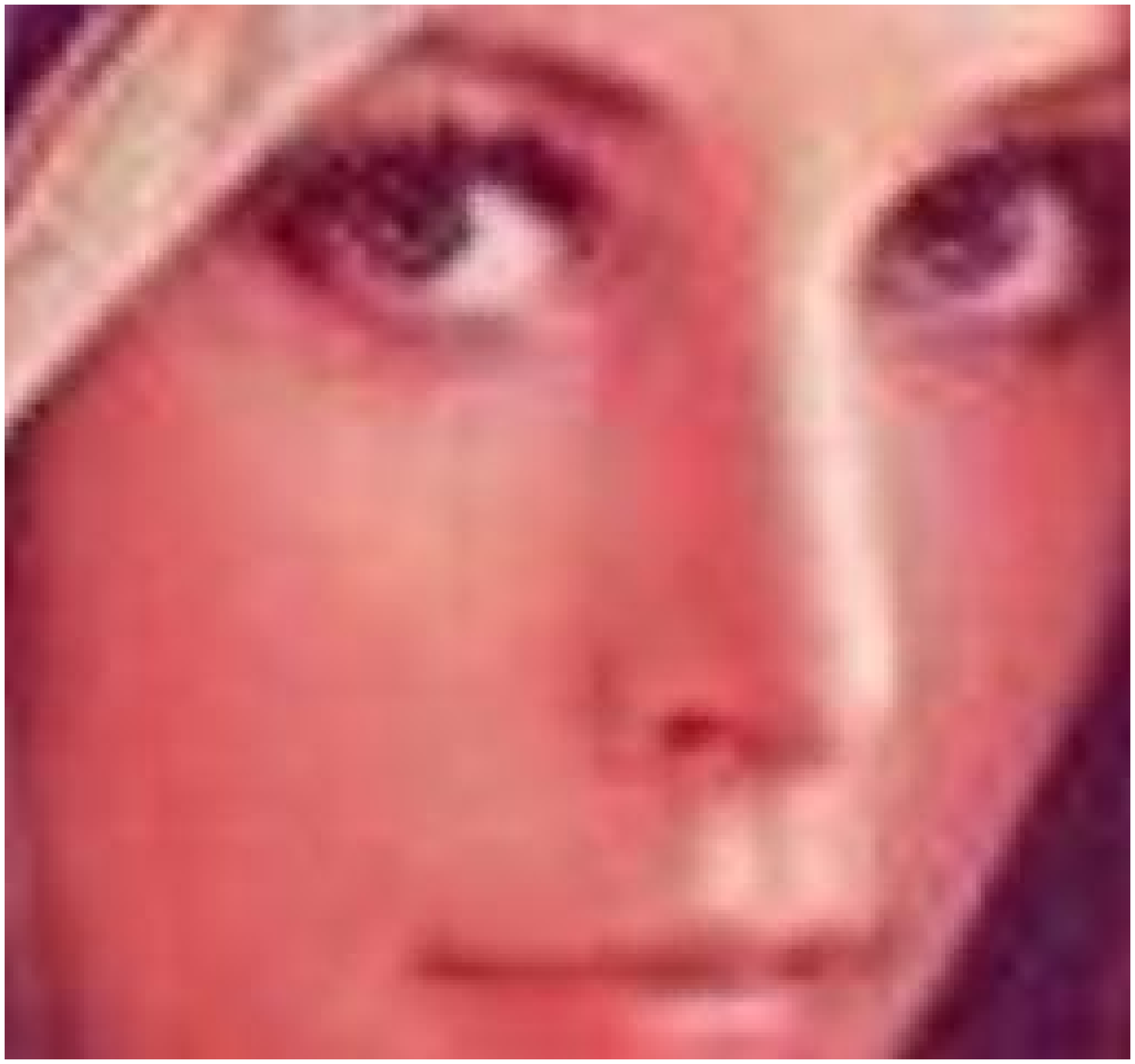}
\includegraphics[width=6.5cm]{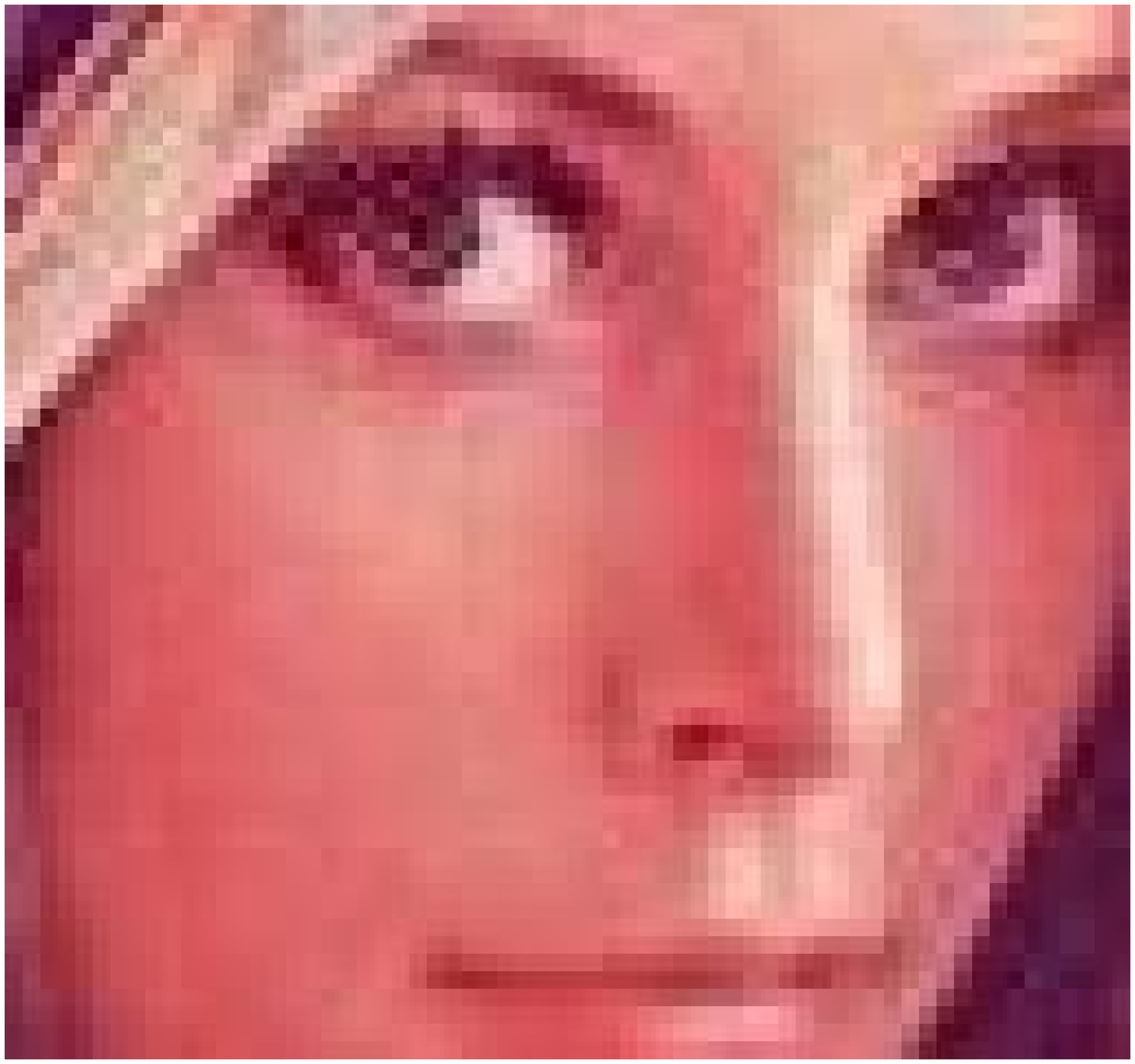}
\end{center}
\caption{Enlargement of the area around the face of Fig.~\ref{lena}
via a) a fractal interpolation function b)
a linear interpolation}\label{lena-face}
\end{figure}
\par\begin{figure}[ht]
\begin{center}
\includegraphics[width=6.5cm]{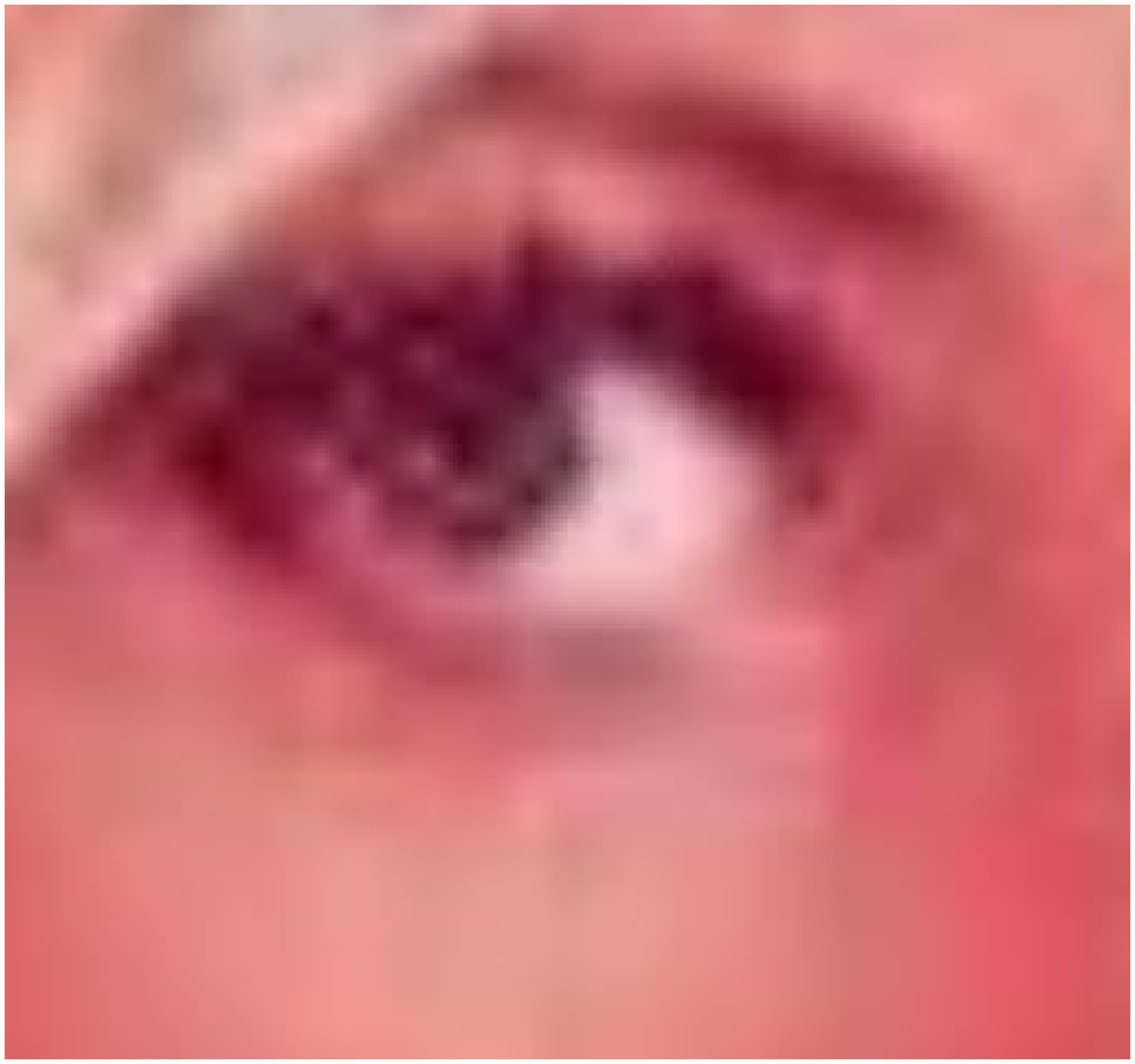}
\includegraphics[width=6.5cm]{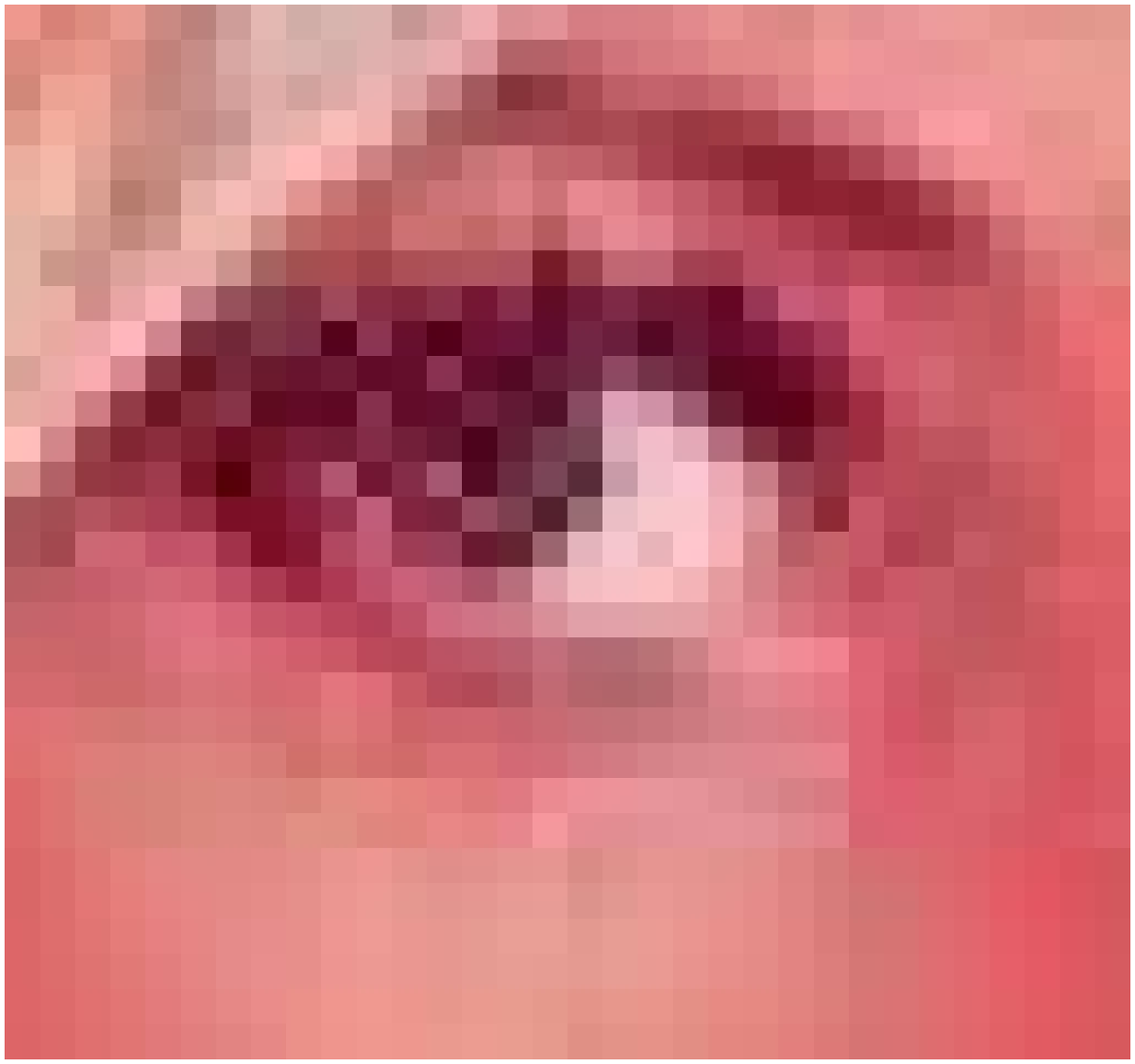}
\end{center}
\caption{Enlargement of the area around the eye of Fig.~\ref{lena}
via a) a fractal interpolation function b)
a linear interpolation}\label{lena-eye}
\end{figure}

\begin{thebibliography}{99}
\bibitem{barn} M.~F.~Barnsley, {\it Fractals Everywhere}
(Academic Press, San Diego, CA, 1993)
\bibitem{ifs} H.~O.~Peitgen, H.~J\"urgens, and D.~Saupe,
{\it Chaos and Fractals -- New Frontiers of Science} (Springer
Verlag, New York, 1992).
\bibitem{sv} H.~O.~Peitgen and D.~Saupe, {\it The Science of
Fractal Images} (Springer Verlag, New York, 1988).
\bibitem{jaq} A.~Jacquin, IEEE trans.~on Image Processing,
January, 1992. 
\bibitem{comp1} M.~Barnsley, L.~Hurd, and L.~Anson,
{\it Fractal Image Compression} (A.~K.~Peters, New York, 1993).
\bibitem{comp2} Y.~Fisher (editor), {\it Fractal Image Compression:
Theory and Application} (Springer Verlag, New York, 1995).
\bibitem{mass} P.~R.~Massopust, {\it Fractal Functions, Fractal 
Surfaces and Wavelets} (Academic Press, San Diego, CA, 1994). 
\end{thebibliography}
\end{document}